\documentclass[preprint,showpacs,preprintnumbers,amsmath,amssymb,nofootinbib,showkeys,aps]{revtex4}
\pdfoutput=1
\usepackage{epsfig}

\usepackage{graphicx}
\usepackage{dcolumn}
\usepackage{bm}
\usepackage{amsmath}
\usepackage{amssymb}
\usepackage{latexsym}
\usepackage{color}
\usepackage{hyperref}
\usepackage{mathrsfs}
\usepackage{float}

\newcommand{\be}{\begin{equation}}
\newcommand{\ee}{\end{equation}}
\newcommand{\bea}{\begin{eqnarray}}
\newcommand{\eea}{\end{eqnarray}}

\newcommand{\like}{\mathscr{L}}
\begin{document}


\title{Dynamical dark energy from Kretschmann scalar at low redshifts}

\author{A. B. Fernandes$^1$}\email{alicia.bassanelli@unesp.br}
\author{S. H. Pereira$^{1}$}\email{s.pereira@unesp.br}
\author{J. F. Jesus$^{2,1}$}\email{jf.jesus@unesp.br}
\author{D. M. Soares Junior$^{1}$}\email{dm.soares@unesp.br}

\affiliation{$^1$Universidade Estadual Paulista (UNESP), Faculdade de Engenharia e Ci\^encias de Guaratinguet\'a, Departamento de F\'isica - Av. Dr. Ariberto Pereira da Cunha 333, 12516-410, Guaratinguet\'a, SP, Brazil\\
$^2$Universidade Estadual Paulista (UNESP), Instituto de Ciências e Engenharia, Departamento de Ci\^encias e Tecnologia - R. Geraldo Alckmin, 519, 18409-010, Itapeva, SP, Brazil
}


\def\zt{\mbox{$z_t$}}

\begin{abstract}
In this work, we present a cosmological model in which the cosmological constant term is replaced by the Kretschmann scalar at the level of the action. In this way, it becomes possible to implement a model of dynamical dark energy. After constraining the free parameters using observational data from supernovae and cosmic chronometers, we show that the model provides a good fit to the observational data. In particular, we show that, at least at low redshifts, the behavior of the equation-of-state parameter $w(z)$ closely reproduces that obtained in phenomenological models that have been recently studied based on the latest observational data from the DESI collaboration. Likewise, the present model also indicates the occurrence of a phantom-crossing regime.
\end{abstract}

\maketitle


\section{Introduction}

Recent results from the Dark Energy Spectroscopic Instrument (DESI) collaboration provide the largest three-dimensional mapping of our Universe to date, offering new insights into the so-called dark energy over a period spanning approximately 11 billion years \cite{DESI:2025fxa,DESI:2025zpo,DESI:2025zgx}. The DESI Data Release 1 includes spectral analyses of more than 18 million objects, including galaxies, quasars, and stars. Although the DESI data alone indicate a preference for the standard $\Lambda$CDM model with a constant cosmological $\Lambda$ term, when combined with other datasets—such as measurements of the cosmic microwave background (CMB), distance measurements from Type Ia supernovae (SNIa), and weak gravitational lensing observations—the preference shifts toward models in which the cosmological component evolves with time, commonly referred to as time-evolving dark energy or dynamical dark energy.

One way to quantify the temporal evolution of the dark energy component is through the so-called equation-of-state parameter 
$w(z)$, defined as the ratio between the pressure and the energy density of the dark component. Some phenomenological parameterizations for 
$w(z)$ were analyzed in \cite{DESI:2025fii} based on observational data from the DESI collaboration, in particular the so-called CPL model (from Chevallier–Polarski–Linder \cite{Chevallier:2000qy,Linder:2002et}), with an expression given by:
\begin{equation}
    w(z) = w_0 + w_a \frac{z}{1+z},\label{eq_wz}
\end{equation}
with $w_0$ and $w_a$ constrained by observations. Such a parametrization is motivated by scalar field models with specific potentials, commonly referred to as quintessence models \cite{Ratra:1987rm,Ferreira:1997hj,Zlatev:1998tr,Ferreira:1997au}. However, these models have the disadvantage of predicting the existence of a new fundamental field or particle in nature, associated with the scalar field $\phi$, which may not be supported from the perspective of the Standard Model of elementary particle physics. In fact, this type of model that invokes a new scalar field is already a key ingredient of modern inflationary models, and we would now be repeating the same procedure to explain the current phase of late-time cosmic acceleration.

Another approach is to explore the possibility that the time-evolving term arises from a modification of the geometric sector of the theory, an idea commonly referred to as modified gravity theories (see \cite{Wetterich:2014bma} and the references therein.). In this context, the Einstein field equations are modified, allowing for generalizations that, in certain cases, can be interpreted as a form of dark energy that varies with time.

In the standard cosmological model, $\Lambda$CDM, the geometric sector of the Friedmann–Lemaître–Robertson–Walker (FLRW) equations arises from the Ricci curvature scalar $R$ appearing in the Einstein–Hilbert action, supplemented by a cosmological constant term $\Lambda$. In the present work, in order to implement a cosmological term that evolves with time, we propose an action in which the cosmological constant is replaced by another scalar invariant of general relativity, namely the Kretschmann scalar $K$. With this simple assumption, we obtain a cosmological model capable of reproducing the behavior given in Eq. (\ref{eq_wz}) for the equation-of-state parameter at low redshift, while also accounting for other quantities indicated by the DESI survey, such as the evolution of the dark energy density, the deceleration parameter and a preference for a phantom crossing phase. 

At this point, it is important to emphasize that gravitational models extending the Einstein–Hilbert action through the inclusion of additional higher-order curvature invariants—such as quadratic terms in the Ricci scalar, the Ricci tensor, and the Riemann tensor, arise naturally within the framework of effective theories of gravity \cite{Nojiri:2003ft}. In particular, these terms can be interpreted as high-energy corrections to General Relativity, becoming relevant in regimes of large curvature, such as the early Universe. From a phenomenological standpoint, such contributions are especially appealing in cosmology, as they can induce phases of accelerated expansion without the need for additional scalar fields, as in quintessence models. A classic example is the $R^2$ Starobinsky model \cite{Starobinsky:1980te}, which provides a successful description of the inflationary phase of the Universe. More generally, quadratic terms can significantly modify the dynamics of the scale factor, allowing not only for inflationary scenarios but also for non-singular solutions or transition regimes between different cosmological eras. However, the inclusion of such terms requires caution \cite{Stelle:1976gc}. In general, quadratic invariants involving the Ricci or Riemann tensors lead to higher-order equations of motion (typically fourth order), which are associated with the emergence of additional degrees of freedom. This may compromise the consistency of the theory as a fundamental description, although it can remain valid as an effective theory within a limited energy range—precisely the regime of interest here, since we focus on the behavior of the model at low redshifts.

Another important aspect is that different forms of the action may lead to equivalent or inequivalent equations of motion, depending on the treatment of total derivative terms. In four dimensions, for instance, specific combinations of curvature invariants are related through the Gauss–Bonnet term, which is topological and does not contribute dynamically. Furthermore, during the derivation of the field equations, integrations by parts may redistribute derivatives among different terms, leading to alternative—but physically equivalent—forms of the equations. Therefore, a careful treatment of boundary terms is essential for a consistent formulation. In the present work, we derived the Friedmann equations both by means of the Euler–Lagrange formalism and through the standard procedure of varying the action with respect to the metric. In both approaches, the same result was obtained, indicating that the boundary terms were properly treated. This ensures that no higher-order derivative terms are discarded through integrations of surface terms.

The article is organized as follows. In Section II, the cosmological equations are derived. In Section III, the analysis and comparative results with the CPL model are presented. Finally, Section IV contains the conclusions. In the Appendix, the Einstein equations are derived using both of the methods mentioned above.

\section{Cosmological equations with Kretschmann scalar}

The action for the model is written as:
\begin{equation}
    S = \int d^4x \sqrt{-g}  \frac{1}{2\kappa}\left(R+\alpha K  \right) +S_m\,,\label{action}
\end{equation}
where $\kappa = 8\pi G$, $R$ is the Ricci curvature scalar, $S_m$ corresponds to the action of the matter sector (barions, radiation and dark matter) and $K$ is the Kretschmann scalar invariant, defined by the contraction of two Riemann tensors, $K=R^{\alpha\beta\gamma\lambda}R_{\alpha\beta\gamma\lambda}$. $\alpha$ is a parameter with dimension of [time]$^2$, which must be constrained by observational data. Notice that the term $\alpha K$ enters the action exactly in the same way as a cosmological constant $\Lambda$ in the standard model. Taking the variation of the first term in the action with respect to the metric would lead to the corresponding Einstein field equations. However, due to the presence of higher-order derivative terms -- arising from the Kretschmann scalar  -- we instead choose to derive the Friedmann equations directly using the Euler–Lagrange formalism. To this end, we write the flat FLRW metric including a lapse function $N(t)$:
\begin{equation}
    ds^2 = -N^2(t)dt^2+a^2(t)[dx^2+dy^2+dz^2]\,.
\end{equation}
Then we write $R$ and $K$ as a function of $N$ and $a$. The Euler–Lagrange equation obtained by varying the Lagrangian density with respect to 
$N(t)$ yields the first Friedmann equation\footnote{Due to the presence of second-order time-derivative terms, we must in fact employ the second-order Euler–Lagrange formalism.}, whereas variation with respect to the scale factor $a(t)$ leads to the second Friedmann equation. At the end of the calculation, we set $N(t)\to 1$. The variation of $S_m$ yields the energy–momentum tensor in the usual way, and we adopt a perfect fluid form for the energy–momentum tensor corresponding to the matter sector.

The corresponding flat Friedmann equations can be written as:
\begin{equation}
    H^2 = \frac{8\pi G}{3} \big[ \rho_m + \rho_K]\,,\label{eqH2}
\end{equation}
\begin{equation}
    3H^2 + 2\dot{H} = -8\pi G \big[P_m + P_K]\,, \label{eqHp}
\end{equation}
where $H\equiv\dot{a}/a$, $\rho_m$ and $P_m$ are the energy density and pressure of standard matter, while $\rho_K$ and $P_K$ are the corresponding ones related to the Kretschmann contribution:
\begin{equation}
    \rho_K = \frac{3}{2\pi G}\alpha \big[H \ddot{H} - \frac{\dot{H}^2}{2} + 3 H^2 \dot{H}]\,,\label{eq_rhok}
\end{equation}
\begin{equation}
    P_K = -\frac{\alpha}{2\pi G}\big[\dddot{H} + 6 H \ddot{H} + 9 H^2 \dot{H} + \frac{9}{2}\dot{H}^2]\,. \label{eq_Pk}
\end{equation}
Written in this form, it is easy to show that the conservation equation is satisfied:
\begin{equation}
    \dot{\rho_K} + 3 H (\rho_K + P_K) = 0\,,
\end{equation}
which guaranties separate conservation for each component. Also, the Kretschmann contribution acts as a dynamical dark energy component.

The first Friedmann equation, together with Eq. (\ref{eq_rhok}), can be solved numerically. For this purpose, it is more convenient to rewrite the equation in terms of the redshift $z$, using the relation $\frac{d}{dt} = -(1+z)H\frac{d}{dz}$. We have:
\begin{equation}
    H^2 + 4\alpha \bigg( 2 (1+z) H^3 H' - \frac{1}{2}(1+z)^2 H^2 H'^2 - (1+z)^2 H^3 H''  \bigg) = H_0^2 \Omega_{m0} (1+z)^3\,,\label{eq_H2z}
\end{equation}
where $H_0$ is the Hubble parameter and we have introduced the matter density parameter $\Omega_{m0}$. Such free parameters must be constrained by observational data, together with $\alpha$ and the initial condition for $H'(0)$. 

It is also convenient to rewrite the energy density and pressure associated with the Kretschmann scalar in terms of the redshift:
\begin{eqnarray}
    \rho_K = \frac{3\alpha}{4\pi G} [2(1+z)^2H^3H'' - 4(1+z)H^3H'+(1+z)^2H^2H'^2]\,,
\end{eqnarray}
\begin{eqnarray}    
    P_K = \frac{\alpha}{4\pi G} [8(1+z)H'H^3 - 13(1+z)^2H'^2 H^2 - 6(1+z)^2H''H^3  \nonumber \\
     -2 (1+z)^3H'^3 H -2(1+z)^3H'''H^3 - 8(1+z)^3H'H''H^2]\,.
\end{eqnarray}

After numerically solving the differential equation (\ref{eq_H2z}) for 
$H(z)$, with free parameters to be constrained by observations, we are interested in evaluating the evolution of several quantities, among them the equation-of-state parameter $w(z)$ associated with the Kretschmann term:
\begin{eqnarray}
    w (z) = \frac{P_K}{\rho_K}\,,\label{wK}
\end{eqnarray}
the energy density $\rho_K$ normalized to its present value:
\begin{eqnarray}
    f (z) = \frac{\rho_K (z)}{\rho_{K0}}\,,\label{fK}
\end{eqnarray}
the deceleration parameter $q(z)$ for the model:
\begin{eqnarray}
    q(z) = - \frac{\ddot{a}\dot{a}}{\dot{a}^2} = (1+z)\frac{H'}{H} -1\,, \label{eq_qz}
\end{eqnarray}
and the $Om(z)$ diagnostic parameter:
\begin{eqnarray}
    Om(z) = \frac{(H(z)/H_0)^2 -1}{(1+z)^3-1}\,.\label{Omz}
\end{eqnarray}
All these quantities should be compared with those presented in Ref. \cite{DESI:2025fii} for the phenomenological model (\ref{eq_wz}) based on the data obtained from the DESI collaboration.

\section{Analysis and results}

\subsection{Initial conditions and free parameters}

In order to solve Eq. (\ref{eq_H2z}) numerically and constrain the free parameters with observational data, we must first express the equation in terms of dimensionless quantities. As discussed previously, the constant $\alpha$ has dimensions of 
[time]$^2$. To work with a dimensionless parameter, we introduce the parameter $\tilde{\alpha}=\alpha H_0^2$, which is indeed dimensionless. Note that the limit $\alpha \to 0$ corresponds to the case in which the contribution from the Kretschmann scalar is absent, which may be favoured when the model is constrained by observational data. Recall that the Kretschmann term enters the model as a substitute for the cosmological constant. In order to avoid numerical instabilities in the limit 
$\alpha \to 0$, we therefore work with the parameter $\gamma = 1/\tilde{\alpha}$.

Having done so, we can rewrite the differential equation (\ref{eq_H2z}) in terms of the dimensionless parameter 
$E(z) = H(z)/H_0$, casting it into the standard form of a second-order differential equation:
\begin{eqnarray}
   E'' = \frac{1}{(1+z)^2 E^3}\bigg[ 2(1+z)E^3E' - \frac{1}{2}(1+z)^2 E^2E'^2 + \frac{\gamma}{4}\bigg(E^2 - \Omega_{m}(1+z)^3\bigg)\bigg]\,.\label{eq_Ez}
\end{eqnarray}
To solve this equation numerically, we require an initial condition for $E'(z)$. From the definition of the deceleration parameter given in Eq. (\ref{eq_qz}), we obtain $H'(0) = (1+q_0)H_0$, where $q_0$ denotes the present-day value of the deceleration parameter. Therefore,  $E'(0)=1+q_0$, and we treat $q_0$ as a free parameter to be constrained by observations. We thus arrive at a set of four free parameters to be determined observationally, namely, $\Omega_m$, $H_0$, $\gamma$ and $q_0$.

\subsection{The observational dataset and methodology}

In order to test the model and constrain its free parameters using observational data, we make use of 32 Hubble parameter data, $H(z)$, compiled by \cite{MorescoEtAl22}, known as cosmic chronometers. This compilation includes statistical and systematic uncertainties described in a covariance matrix. Additionally, we use the Pantheon+\&SH0ES sample \cite{pantheon+}, consisting of 1701 light curves for 1550 distinct SNe Ia in the redshift range $0.001 < z < 2.26$. This sample includes the SH0ES Cepheid host distances \cite{sh0es}, used to calibrate the SNe Ia magnitudes. 

The values of the free parameters were obtained by using Bayesian statistics, with a flat prior over the parameters and a likelihood of the form $\like \propto e^{-\frac{1}{2}\chi ^2}$. The affine invariant method of Monte Carlo Markov chain analysis (MCMC) was used with {\sffamily emcee} software \cite{GoodmanWeare,ForemanMackey13}, implemented in {\sffamily Python} language. 

The range used for the parameters were: $-19.5 < M < -19.0$, $30<H_0\text{ (km/s/Mpc) }<100$, $0<\Omega_m <1$, $-80 < \gamma < 80$ and $-5 < q_0 < 5$. We have also used an additional prior over $\Omega_m$ from the KiDS-1000 survey \cite{Kids}, namely $\Omega_m=0.270\pm0.079$, which is a symmetrization of the KiDS result, $\Omega_m=0.270^{+0.056}_{-0.102}$, according to D'Agostini symmetrization rules \cite{DAgostini}.

\subsection{Results}

\begin{figure}[ht]
    \centering
    \includegraphics[width=.9\textwidth]{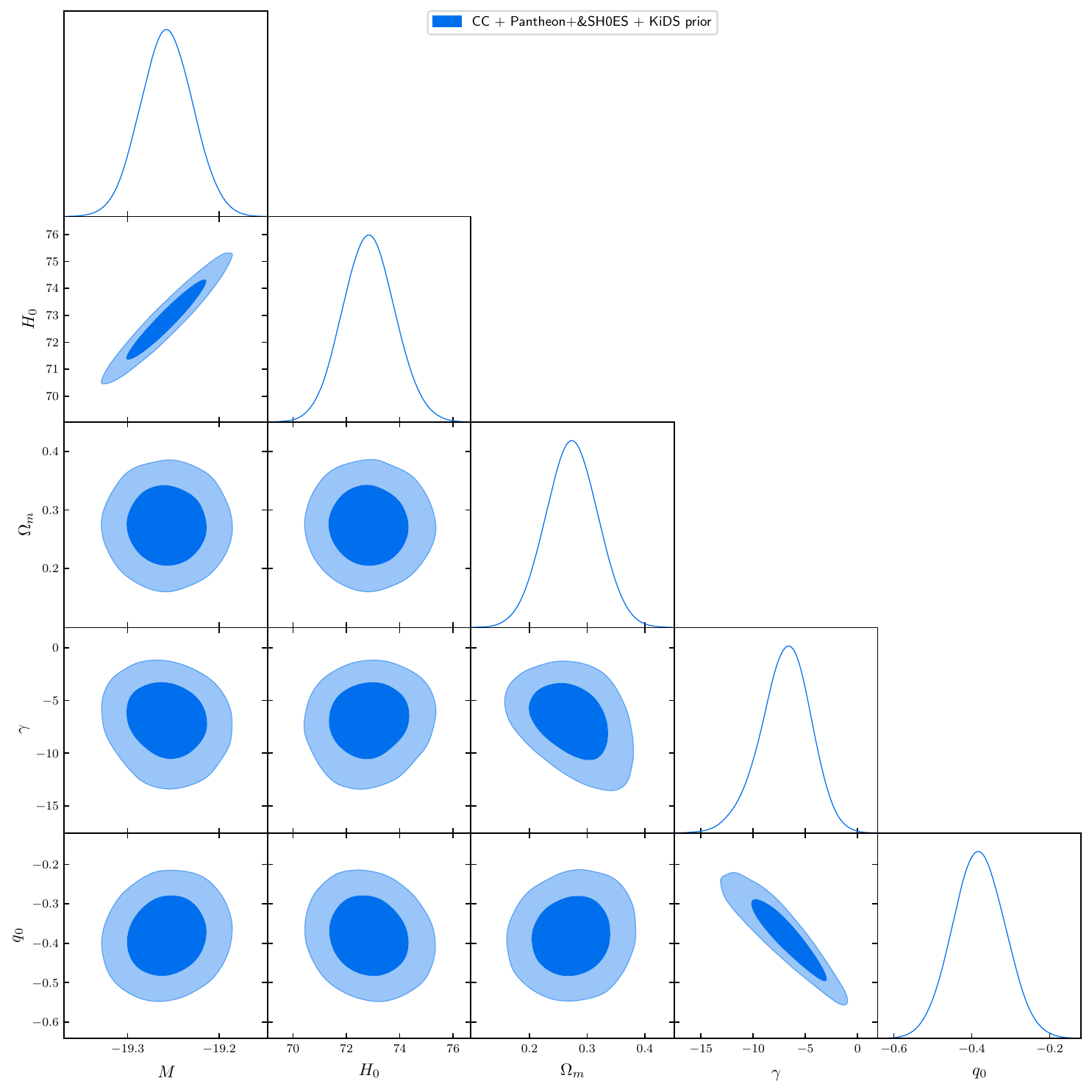}
    \caption{Contours for the joint analysis of $H(z)$ data from CC, SNe Ia from Pantheon+\&SH0ES and KiDS prior at 1$\sigma$ and 2$\sigma$ for the free parameters.}
    \label{figure1}
\end{figure}

The results for the free parameters with the joint analysis of $H(z)$ data from CC, SNe Ia from Pantheon+\&SH0ES and KiDS prior at 1$\sigma$ and 2$\sigma$ are presented in Figure \ref{figure1}. The values at 68\% confidence level (c.l.) are in Table I.

\begin{table}[ht]
    \centering
\begin{tabular} { l c}
\centering
 Parameter &  68\% limits\\
\hline
{\boldmath$M              $} & $-19.257\pm 0.029 $\\

{\boldmath$H_0            $} & $72.85\pm 0.98        $\\

{\boldmath$\Omega_m       $} & $0.273\pm0.045   $\\

{\boldmath$\gamma         $} & $-6.9^{+2.6}_{-2.2}        $\\

{\boldmath$q_0            $} & $-0.382\pm 0.067     $\\
\hline \label{tab01}
\end{tabular}
\caption{Mean values and 68\% c.l. constraints for the parameters $M$, $H_0$, $\Omega_m$, $\gamma$ and $q_0$.}
\end{table}

It is evident that the values of $H_0$ and $\Omega_m$ are consistent with those of the standard model. In particular, it is evident that the obtained value of $H_0$ is highly consistent with local measurements, in agreement with the results reported by the SH0ES collaboration. The values of $q_0$ and $\gamma$ are also presented, being parameters specific to the model, with $q_0$ corresponding to an initial condition and $\gamma$ quantifying the coupling of the Kretschmann scalar through $\alpha = 1/\gamma H_0^2$. As will be shown below, the model is highly sensitive to small variations in $q_0$ and $\gamma$ which are responsible for generating a dynamical dark energy component in agreement with the DESI results.

\subsection{Comparison with DESI results}

In order to compare our model with that represented by the phenomenological choice of the equation-of-state parameter  $w(z)$ given by (\ref{eq_wz}) in the DESI article \cite{DESI:2025fii}, we construct the corresponding plots of $w(z)$, $q(z)$, $f(z)$ and $Om(z)$, given by Eqs. (\ref{wK}) - (\ref{Omz}). We will refer to the DESI results simply as $w_0 w_a CDM$ (DESI), or DESI model, and our model as Kretschmann.

\begin{figure}[t]
    \centering
    \includegraphics[width=.50\textwidth]{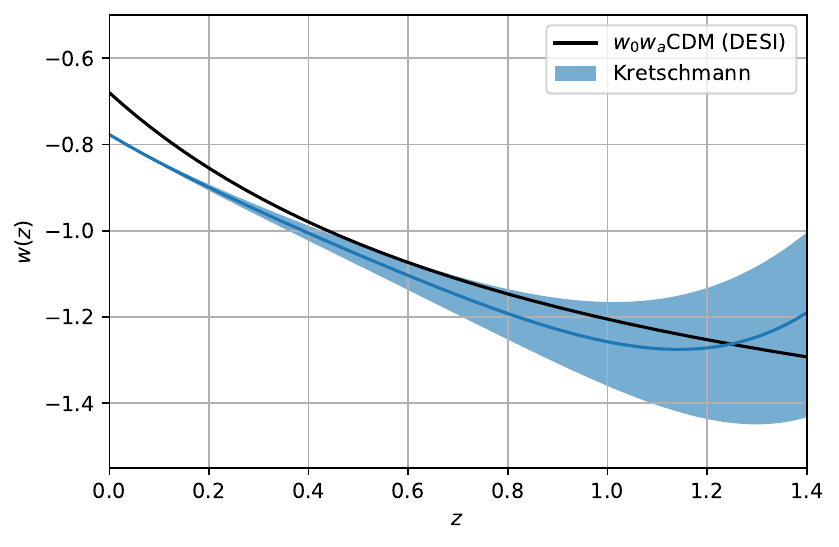}
    \includegraphics[width=.49\textwidth]{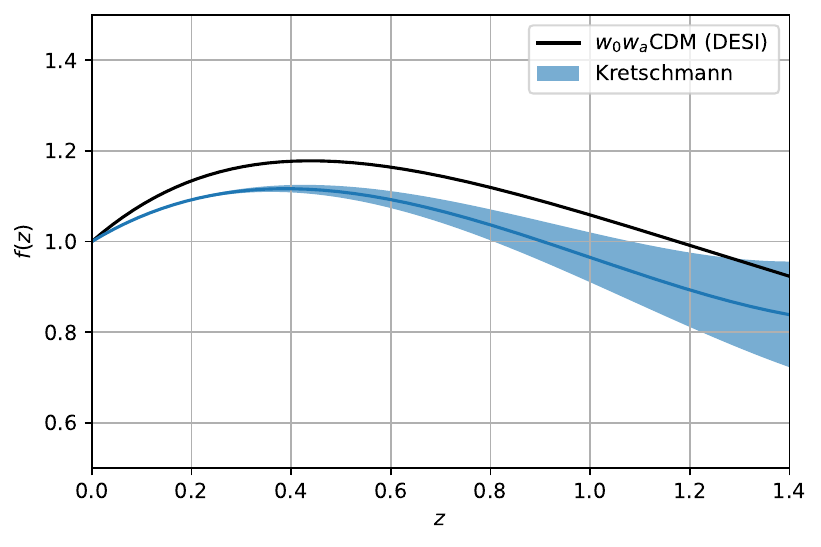}
    \includegraphics[width=.50\textwidth]{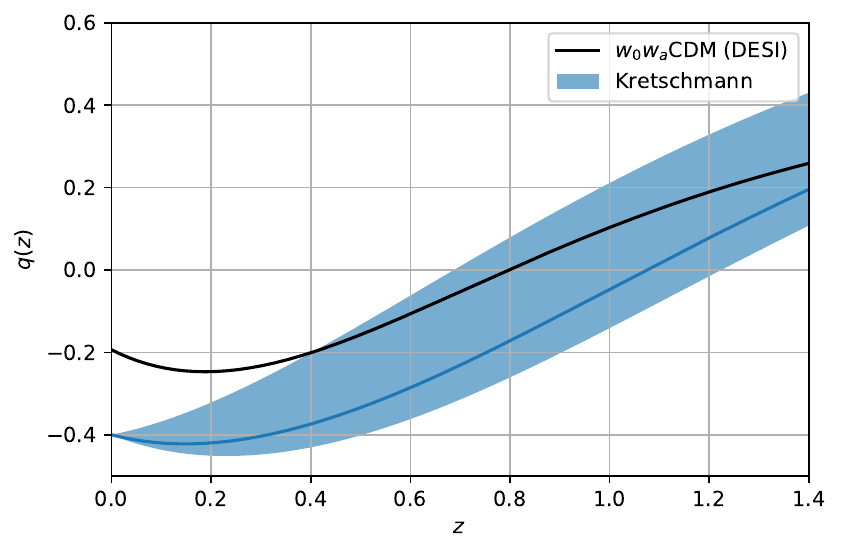}
    \includegraphics[width=.49\textwidth]{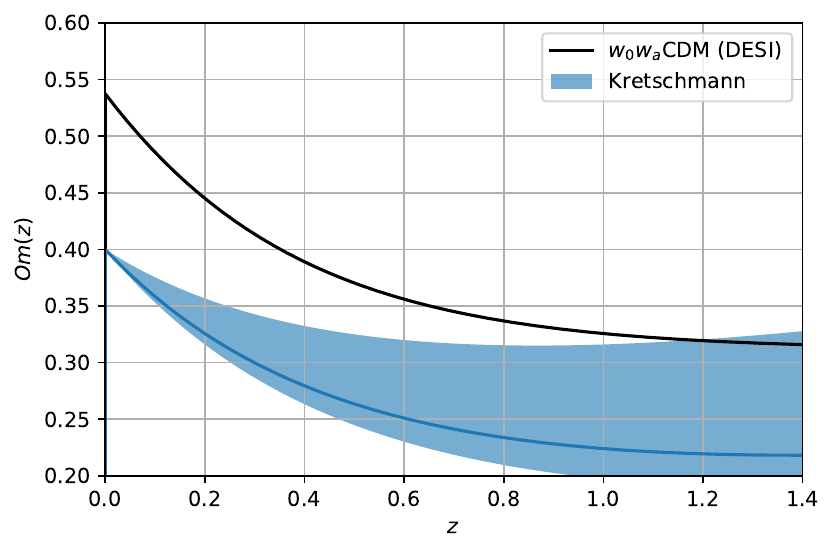}
    \caption{Comparison between the results obtained for the DESI model and our Kretschmann model by varying $\gamma$. For the DESI model, the values of $w_0 = -0.68$ and $w_a = -1.05$ were used. For our model, we employed $H_0 = 72.8$ km/s/Mpc, $\Omega_m = 0.228$, $q_0 = -0.4$ and $\gamma = -8.2$. The blue shaded region was constructed using the interval $\gamma = [-8.4, -8.0]$ for $w(z)$ and $f(z)$ and $\gamma = [-9.1, -5.3]$ for $q(z)$ and $Om(z)$.}
    \label{figure2}
\end{figure}

Figure \ref{figure2} presents the main results. The redshift evolution of the parameters obtained in the DESI collaboration analysis (shown in black line in the plots) is compared with those of our model (shown in blue). As described in the figure captions, we use parameter values within the 1 $\sigma$ range to generate the plots, not necessarily the central values, since a better agreement with the DESI model is achieved for slightly different values that still lie within the 1 $\sigma$ interval. 

We find that the curves of our model are highly sensitive to small variations in 
$q_0$ and $\gamma$, as evidenced by the shaded contours constructed over different intervals of $\gamma$. At low redshifts, the behavior of $w(z)$ and $f(z)$ closely matches that obtained by DESI model. For 
$q(z)$ and $Om(z)$, the overall behavior is likewise preserved. The behavior of the parameter of equation of state $w(z)$ is particularly noteworthy. Our model also predicts a crossing of the phantom divide at $z$ near $0.40$, very close to the DESI result of $z \approx 0.45$. The normalized dark energy density $f(z)$ associated with our model also follows the same behavior as that of the DESI model. A maximum is reached around  $z\approx 0.45$, followed by a subsequent decrease. For the deceleration parameter $q(z)$, however, the initial behavior is significantly different, due to our choice of treating $q_0$ as a free parameter to be constrained by the model. While in the DESI model the value 
$q_0=-0.2$, in our model the best fits are obtained for $q_0=-0.4$. Nevertheless, the possibility of a transition redshift around 
$z\approx0.8$ is still observed within the range considered for $\gamma$. Finally, the diagnostic function $Om(z)$ exhibits the same qualitative behavior, although with a clear downward shift relative to that of the DESI model.

\begin{figure}[t]
    \centering
    \includegraphics[width=.50\textwidth]{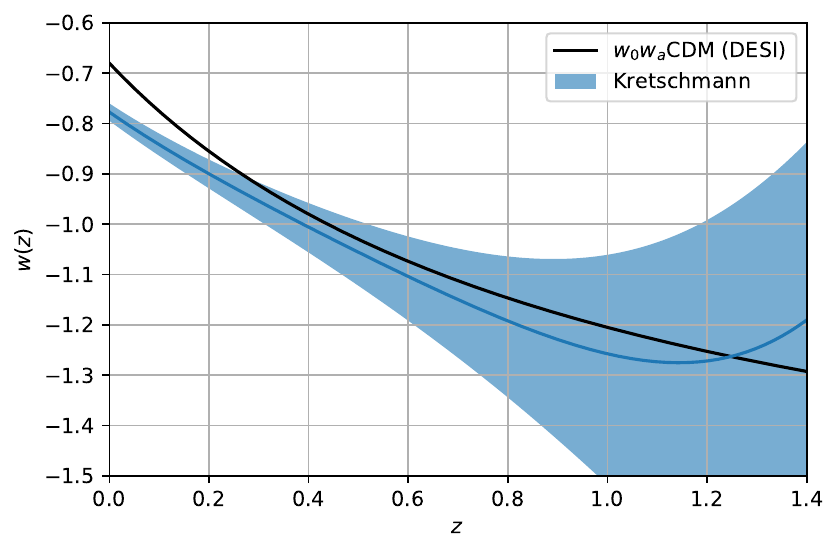}
    \includegraphics[width=.49\textwidth]{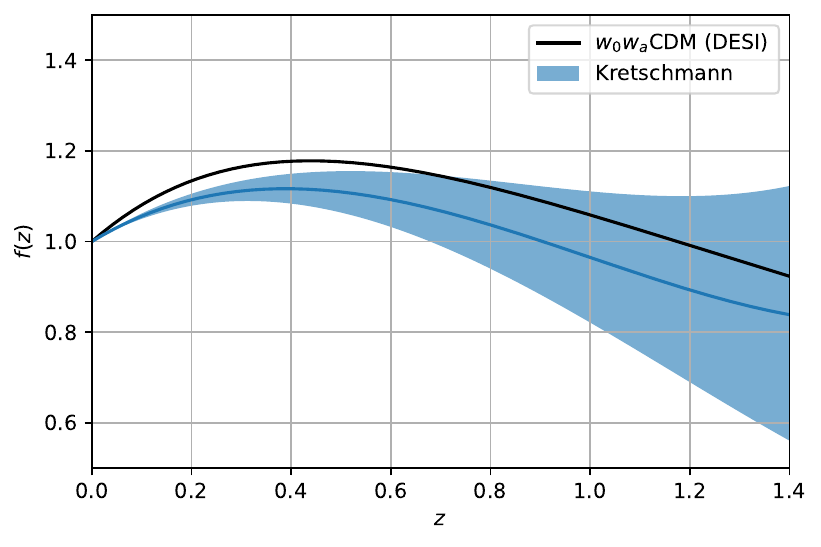}
    \includegraphics[width=.50\textwidth]{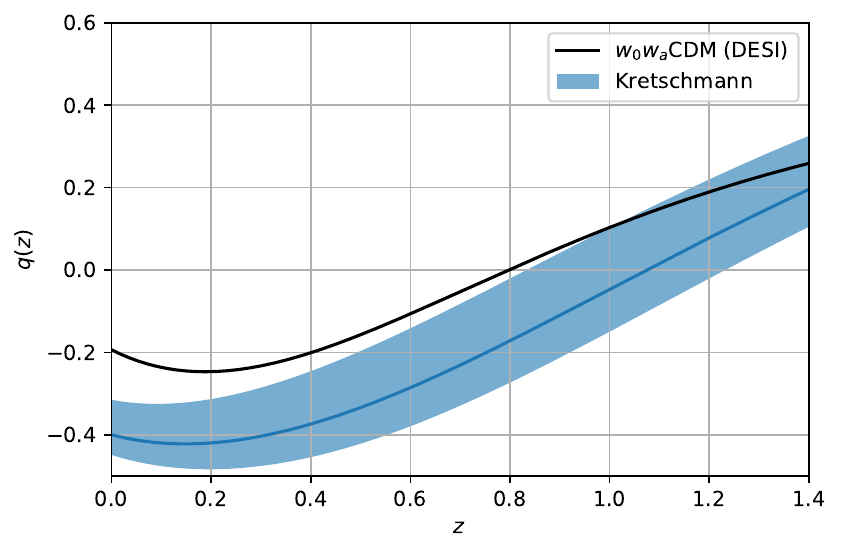}
    \includegraphics[width=.49\textwidth]{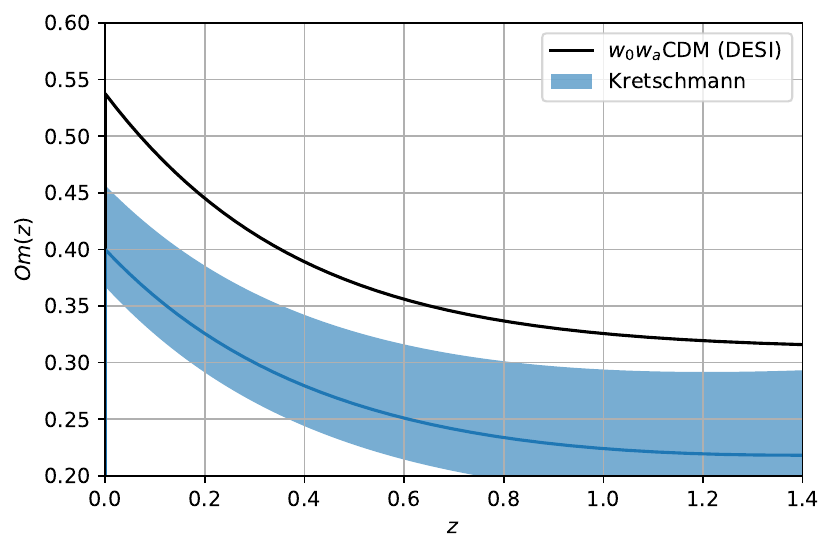}
    \caption{Comparison between the results obtained for the DESI model and our Kretschmann model by varying $q_0$. For the DESI model, the values of $w_0 = -0.68$ and $w_a = -1.05$ were used. For our model, we employed $H_0 = 72.8$ km/s/Mpc, $\Omega_m = 0.228$, $q_0 = -0.4$ and $\gamma = -8.2$. The blue shaded region was constructed using the interval $q_0 = [-0.42, -0.38]$ for $w(z)$ and $f(z)$ and $q_0 = [-0.449, -0.315]$ for $q(z)$ and $Om(z)$.}
    \label{figure3}
\end{figure} 

In Figure \ref{figure3}, the same comparative analysis is performed, now allowing for variations in the initial condition $q_0$, still within the $1\sigma$ range. The same qualitative features are observed, with particular emphasis on a larger variation of $w(z)$ at higher redshifts.

Both analyses show that, at low redshifts, the model is capable of reproducing a dynamical dark energy component, with a time-varying equation-of-state parameter, in agreement with the results obtained by DESI. This opens up the possibility that the behavior of dynamical dark energy may be, at least in part, a consequence of modifications to the geometric sector of the theory, rather than necessarily requiring the existence of a new material component.
\newpage

\section{Conclusion}

In this work, we have analyzed a cosmological model in which the cosmological constant is replaced by the Kretschmann scalar at the level of the action, characterized by a coupling constant $\alpha$. After constraining the free parameters of the model with observational data, we obtained good agreement for the parameters $H_0$ and $\Omega_m$, in addition to identifying a dynamical dark energy component driven by the Kretschmann scalar. By adopting best-fit values within the 1 $\sigma$ range, we were able to reproduce the time-varying equation of state suggested phenomenologically by the DESI collaboration analysis. The existence of a crossing of the phantom divide was also verified.

In contrast to scalar field and quintessence models, which are commonly employed to describe dynamical dark energy, we have demonstrated the viability of attributing this behavior to a geometric origin, namely the inclusion of a Kretschmann scalar term in the action. At low redshifts, the resulting behavior is very similar to that of the phenomenological $w_0 w_a$CDM model. Although the behavior of the equation-of-state parameter differs at higher redshifts, the present model proves to be a robust effective description in the low-redshift regime, without the need to introduce additional scalar fields into the theory. The possibility of including other curvature invariants to account for high-redshift effects is an interesting direction that will be explored in future work.

\appendix
\section{The Friedmann equations from Euler-Lagrange formalism}
The Friedmann equations associated with the modified gravitational action with the Kretschmann scalar, $K$, are obtained from the Lagrangian formulation using the Euler–Lagrange equations for Lagrangians with higher-order derivatives \cite{Hobson:2006se}:
\begin{align}
\frac{d^2}{dt^2} \left(\frac{\partial\mathcal{L}}{\partial \ddot{q}_i}\right) - \frac{d}{dt} \left(\frac{\partial \mathcal{L}}{\partial \dot{q}_i}\right) + \frac{\partial \mathcal{L}}{\partial q_i} = 0 \, , \label{euler geral}
\end{align}
where $\mathcal{L}=\mathcal{L}(q_i,\dot{q}_i,\ddot{q}_i)$ is the Lagrangian density and $q_i$ denotes the gravitational degrees of freedom.

Considering the flat FLRW metric, to correctly obtain both Friedmann equations, we adopt the general form of the metric with the lapse function $N(t)$ parameterizing the temporal component:
\begin{equation}
ds^2 = -N^2(t)\,dt^2 + a^2(t) [dx^2+dy^2+dz^2] \,. \label{metrica}
\end{equation}
We postpone fixing the gauge $N(t)=1$, which will be imposed only at the end of the analysis.

The modified gravitational action is given by
\begin{equation}
S = \int d^4x \sqrt{-g}\, \frac{1}{2\kappa}\,(R + \alpha K)\,, \label{acao}
\end{equation}
where $g$ is the determinant of the metric, $\kappa=8\pi G$, $R$ is the Ricci scalar, and $\alpha$ is a proportionality constant. The scalars $R$ and $K$ in the action are built from the contractions of the Riemann tensor:
\begin{align}
{R^{\rho}}_{\lambda\mu\nu} = \partial_{\mu} {\Gamma^{\rho}}_{\nu\lambda} - \partial_{\nu} {\Gamma^{\rho}}_{\mu\lambda} + {\Gamma^{\eta}}_{\nu\lambda}{\Gamma^{\rho}}_{\mu\eta} - {\Gamma^{\eta}}_{\mu\lambda} {\Gamma^{\rho}}_{\nu\eta}\,.
\end{align}
For the metric given in Eq. \eqref{metrica}, the non-vanishing components of the Levi-Civita connection are:
\begin{align}
{\Gamma^{0}}_{00} &= \frac{\dot{N}}{N}\,, \\ {\Gamma^{0}}_{ij} = {\Gamma^{0}}_{ji} &= \frac{\dot{a}a}{N^2}\delta_{ij}\,,  \\  {\Gamma^{i}}_{0j}={\Gamma^{i}}_{j0} &= \frac{\dot{a}}{a} {\delta^{i}}_{j}\,.
\end{align}
The non-vanishing components of the Riemann tensor are:
\begin{align}
R^{0}{}_{i0j} = - R^{0}{}_{ij0} &= \left(\frac{\ddot{a}a}{N^2} - \frac{\dot{a}a\dot{N}}{N^3} \right) \delta_{ij}\,, \\
R^{i}{}_{00j} = - R^{i}{}_{0j0} &= \left( \frac{\ddot{a}}{a} - \frac{\dot{a}\dot{N}}{aN} \right){\delta^{i}}_{j}\,, \\
R^{i}{}_{jlm} &= \frac{\dot{a}^2}{N^2}(\delta_{mj}\,{\delta^{i}}_{l} - \delta_{jl}\,{\delta^{i}}_{m})\,,
\end{align}
while the corresponding Ricci tensor components are:
\begin{equation}
R_{00}=\frac{3\dot{a}\dot{N}}{aN}-3\frac{\ddot a}{a},
\qquad
R_{0i}=0,
\qquad
R_{ij}=\left(\frac{a \ddot a}{N^2}+2\frac{\dot a^2}{N^2} - \frac{a \dot{a}\dot{N}}{N^3}\right)\delta_{ij},
\label{eq:appB_ricci_components}
\end{equation}
Notice that all these components also depend on the time derivatives of $N(t)$ and not only on the derivatives of the scale factor.

The Ricci scalar is obtained from the contraction of the Ricci tensor:
\begin{align}
R = g^{\mu\nu} R_{\mu\nu}\,,
\end{align}
where $R_{\mu\nu}= {R^{\rho}}_{\mu\rho\nu}$. Hence, we obtain:
\begin{align}
R = 6\left(\frac{\ddot{a}}{aN^2} - \frac{\dot{a}\dot{N}}{aN^3} + \frac{\dot{a}^2}{a^2N^2} \right)\,.
\end{align}

The Kretschmann scalar is defined by the contraction:
\begin{equation}
K = R^{\rho}{}_{\lambda\mu\nu} \, R_{\rho}{}^{\lambda\mu\nu}\,.
\end{equation}
Then,
\begin{align}
K = 12\left( \frac{\dot{a}^2\dot{N}^2}{a^2N^6} - 2\frac{\ddot{a}\dot{a}\dot{N}}{a^2N^5} + \frac{\ddot{a}^2}{a^2N^4} + \frac{\dot{a}^4}{a^4N^4} \right)\,.
\end{align}
Thus, we can construct the gravitational Lagrangian density from the action given in Eq. \eqref{acao}, defined as:
\begin{align}
\mathcal{L}_g &= \sqrt{-g} \frac{1}{2\kappa}\, (R + \alpha K)\, \nonumber \\
&= \frac{3}{\kappa}\left[ \frac{\ddot{a}a^2}{N} - \frac{\dot{a}a^2\dot{N}}{N^2} +\frac{\dot{a}^2a}{N} + 2\alpha \left( \frac{\dot{a}^2a\dot{N}^2}{N^5} - 2\frac{\ddot{a}\dot{a}a\dot{N}}{N^4} + \frac{\ddot{a}^2a}{N^3} + \frac{\dot{a}^4}{aN^3}\right) \right] \, . \label{Lg}
\end{align}
The obtained form of the Lagrangian makes evident its dependence on higher-order derivatives. Although some terms in \eqref{Lg} can be rewritten as total derivatives (boundary terms), it is preferable to keep them in order to preserve the complete structure of the Lagrangian during the variation process until the gauge $N(t)=1$ is fixed. Thus, even in the presence of higher-order derivatives, the boundary terms do not contribute to the equations of motion after gauge fixing.

Therefore, the equations of motion of the system must be obtained from the general Euler–Lagrange equation, Eq. \eqref{euler geral}, applied to the degrees of freedom $a(t)$ and $N(t)$, for the Lagrangian density:
\begin{align}
\mathcal{L} (a, \dot{a}, \ddot{a}, N, \dot{N}) &= \mathcal{L}_g \label{L total}\,.
\end{align}
In this Appendix, we derive the vacuum equations $(\mathcal{L}_m=0)$ for simplicity. The inclusion of matter is straightforward and is discussed in the main article. 

\subsection{First Friedmann Equation}

The first Friedmann equation is obtained from the Euler–Lagrange equation for $q_i=N(t)$. However, since the Lagrangian density does not depend on $\ddot{N}$, the general expression \eqref{euler geral} reduces to:
\begin{align}
- \frac{d}{dt} \left(\frac{\partial \mathcal{L}}{\partial \dot{N}}\right) + \frac{\partial \mathcal{L}}{\partial N} = 0 \,. \label{Euler_N}
\end{align}

We compute the derivatives of the Lagrangian and impose the gauge condition $N(t)=1$, which implies $\dot{N}=0$ and $\ddot{N}=0$. This yields:
\begin{subequations} \label{eq:derivadas_N calibre}
\begin{align}
\frac{\partial \mathcal{L}}{\partial N}\bigg|_{N=1} &= -\frac{3}{\kappa} \left[\ddot{a}a^2 + \dot{a}^2a + 6\alpha \left(\ddot{a}^2a + \frac{\dot{a}^4}{a}\right)\right]\,, \\
\frac{d}{dt} \left(\frac{\partial \mathcal{L}}{\partial \dot{N}}\right)\bigg|_{N=1} &= -\frac{3}{\kappa} \left[ \ddot{a}a^2 + 2\dot{a}^2a + 2\alpha \left( 2\dddot{a} \,\dot{a}a + 2\ddot{a}^2a + 2\ddot{a}\dot{a}^2 \right)\right] \,.
\end{align}
\end{subequations}
Substituting these expressions into Eq. \eqref{Euler_N}, we obtain the first modified Friedmann equation:
\begin{align}
\frac{\dot{a}^2}{a^2} - 2\alpha\left(\frac{\ddot{a}^2}{a^2}- 2\frac{\ddot{a}\dot{a}^2}{a^3}  -2\frac{\dddot{a} \,\dot{a}}{a^2}  + 3\frac{\dot{a}^4}{a^4} \right) &= 0 \,. \label{eq: Friedmann 1}
\end{align}

\subsection{Second Friedmann Equation}

The second modified Friedmann equation is obtained analogously to the first, now considering $q_i = a(t)$ in Eq. \eqref{euler geral}, which takes the form:
\begin{align}
\frac{d^2}{dt^2} \left(\frac{\partial\mathcal{L}}{\partial \ddot{a}}\right) - \frac{d}{dt} \left(\frac{\partial \mathcal{L}}{\partial \dot{a}}\right) + \frac{\partial \mathcal{L}}{\partial a} = 0 \,. \label{eq: euler_a}
\end{align}
Since the Lagrangian depends explicitly on $\ddot{a}$, the general form of the Euler–Lagrange equation must be used.

We then compute the necessary time derivatives for the application of the Euler–Lagrange equation, evaluated after imposing the gauge condition:
\begin{subequations}
\begin{align}
\frac{\partial \mathcal{L}}{\partial a}\bigg|_{N=1} &= \frac{6}{\kappa}\left[ 2\ddot{a}a+\dot{a}^2 + 2\alpha\left( \ddot{a}^2 -\frac{\dot{a}^4}{a^2} \right) \right] \,, \\
\frac{d}{dt} \left(\frac{\partial \mathcal{L}}{\partial \dot{a}}\right)\bigg|_{N=1} &= \frac{6}{\kappa} \left[ \ddot{a}a +\dot{a}^2 + 4\alpha\left(3\frac{\ddot{a}\dot{a}}{a} -\frac{\dot{a}^4}{a^2} \right)\right]\, , \\
\frac{d^2}{dt^2} \left(\frac{\partial \mathcal{L}}{\partial \ddot{a}}\right)\bigg|_{N=1} &= \frac{6}{\kappa} \left[\ddot{a}a + \dot{a}^2 +2\alpha\left(\ddot{a}^2 + \dddot{a} \,\dot{a} +\ddddot{a} \,a \right) \right]\,,
\end{align}
\end{subequations}
where $a^{(4)}$ denotes the fourth time derivative of the scale factor. Substituting these results into \eqref{eq: euler_a}, we obtain the second modified Friedmann equation in terms of the scale factor:
\begin{align}
\frac{\dot{a}^2}{a^2} + 2\frac{\ddot{a}}{a} + 2\alpha\left(3\frac{\ddot{a}^2}{a^2}+4\frac{\dddot{a} \,\dot{a}}{a^2} +2\frac{\ddddot{a} \,}{a} - 12\frac{\ddot{a}\dot{a}^2}{a^3} + 3\frac{\dot{a}^4}{a^4} \right) &= 0\,. \label{eq: Friedmann 2}
\end{align}

Thus, equations \eqref{eq: Friedmann 1} and \eqref{eq: Friedmann 2} represent the two Friedmann equations for the modified gravity model with the Kretschmann scalar, obtained through the Euler–Lagrange method for Lagrangians with higher-order derivatives. Note that the terms proportional to $\alpha$, originating from the scalar $K$, introduce contributions up to third order in the derivatives of $H$, which correspond to corrections to the cosmological dynamics relative to the standard $\Lambda$CDM model.

\section{Covariant derivation of the Friedmann equations}
\label{app:covariant_friedmann1}

Here we rederive in details the cosmological equations obtained in Appendix A by performing the metric variation of the action \eqref{action}. Since the purpose here is only to provide an independent consistency check of the previous result, we keep the derivation at the covariant level until the last step and only then specialize to the flat FLRW metric \eqref{metrica} in the cosmic-time gauge, \(N=1\).

We start from the variation of the Ricci scalar, which is presented in several textbooks \cite{Hobson:2006se}:
\begin{equation}
    \delta(\sqrt{-g}R) = \delta(\sqrt{-g}g^{\mu\nu}R_{\mu\nu})= (R_{\mu\nu} - \frac{1}{2}g_{\mu\nu}R)\delta{g^{\mu\nu}}
\end{equation}
which defines the Einstein tensor:
\begin{equation}
    G_{\mu\nu} = R_{\mu\nu} - \frac{1}{2}g_{\mu\nu}R\,.
\end{equation}

For the Kretschmann scalar (A14), we have the variation:
\begin{align}
\delta\!\left(R_{\mu\nu\alpha\beta}R^{\mu\nu\alpha\beta}\right)
&=
R_{\mu\nu\alpha\beta}\,\delta R^{\mu\nu\alpha\beta}
+
R^{\mu\nu\alpha\beta}\,\delta R_{\mu\nu\alpha\beta}
\nonumber\\
&=
R_{\mu\nu\alpha\beta}\,
\delta\!\left(g^{\mu \lambda}g^{\nu \gamma}g^{\alpha \rho}g^{\beta \sigma}R_{\lambda\gamma\rho\sigma}\right)
+
R^{\mu\nu\alpha\beta}\,\delta R_{\mu\nu\alpha\beta}
\nonumber\\
&=
R_{\mu\nu\alpha\beta}
\Big(
g^{\nu \gamma}g^{\alpha \rho}g^{\beta \sigma}R_{\lambda\gamma\rho\sigma}\,\delta g^{\mu \lambda}
+
g^{\mu \lambda}g^{\alpha \rho}g^{\beta \sigma}R_{\lambda\gamma\rho\sigma}\,\delta g^{\nu \gamma}
\nonumber\\
&\quad+
g^{\mu \lambda}g^{\nu \gamma}g^{\beta \sigma}R_{\lambda\gamma\rho\sigma}\,\delta g^{\alpha \rho}
+
g^{\mu \lambda}g^{\nu \gamma}g^{\alpha \rho}R_{\lambda\gamma\rho\sigma}\,\delta g^{\beta \sigma} \nonumber\\
&\quad+
g^{\mu \lambda}g^{\nu \gamma}g^{\alpha \rho}g^{\beta \sigma}\,\delta R_{\lambda\gamma\rho\sigma}
\Big)+
R^{\mu\nu\alpha\beta}\,\delta R_{\mu\nu\alpha\beta}
\nonumber\\
&=
2R^{\mu\nu\alpha\beta}\,\delta R_{\mu\nu\alpha\beta}
+
R_{\mu\nu\alpha\beta}
\Big(
R_{\lambda}{}^{\nu\alpha\beta}\,\delta g^{\mu \lambda}
+
R^{\mu}{}_{\gamma}{}^{\alpha\beta}\,\delta g^{\nu \gamma}
+
R^{\mu\nu}{}_{\rho}{}^{\beta}\,\delta g^{\alpha \rho}
+
R^{\mu\nu\alpha}{}_{\sigma}\,\delta g^{\beta \sigma}
\Big)
\nonumber\\
&=
2R^{\mu\nu\alpha\beta}\,\delta R_{\mu\nu\alpha\beta}
+
4R_{\mu}{}^{\alpha\beta\gamma}R_{\nu\alpha\beta\gamma}\,\delta g^{\mu\nu}
\nonumber\\
&=
2R^{\mu\nu\alpha\beta}\,\delta (g_{\mu\gamma} R^{\gamma}{}_{\nu\alpha\beta})
+
4R_{\mu}{}^{\alpha\beta\gamma}R_{\nu\alpha\beta\gamma}\,\delta g^{\mu\nu}
\nonumber\\
&=
2R^{\sigma\nu\alpha\beta}(\delta g_{\sigma\gamma} R^{\gamma}{}_{\nu\alpha\beta} + g_{\sigma\gamma} \delta R^{\gamma}{}_{\nu\alpha\beta})
+
4R_{\mu}{}^{\alpha\beta\gamma}R_{\nu\alpha\beta\gamma}\,\delta g^{\mu\nu}
\nonumber\\
&=
2R_{\mu}{}^{\alpha\beta\gamma}\,\delta R^{\mu}{}_{\alpha\beta\gamma}
+
2R_{\mu}{}^{\alpha\beta\gamma}R_{\nu\alpha\beta\gamma}\,\delta g^{\mu\nu},
\label{eq:BdeltaKraw}
\end{align}
where we have made use of the identity $\delta g_{\sigma\gamma} = -g_{\mu\sigma}g_{\nu\gamma}\delta g^{\mu\nu}$. Therefore,
\begin{equation}
\delta\!\left(\sqrt{-g}\,K\right)
=
\sqrt{-g}
\left[
-\frac{1}{2}g_{\mu\nu}K
+
2R_{\mu}{}^{\alpha\beta\gamma}R_{\nu\alpha\beta\gamma}
\right]\delta g^{\mu\nu}
+
2\sqrt{-g}\,R_{\mu}{}^{\alpha\beta\gamma}\,\delta R^{\mu}{}_{\alpha\beta\gamma},
\label{eq:BdeltaSqrtgK}
\end{equation}
where \(\delta\sqrt{-g}=-\frac12\sqrt{-g}\,g_{\mu\nu}\delta g^{\mu\nu}\).

The last term is treated with the Palatini identity, 
\begin{equation}
\delta R^{\mu}{}_{\alpha\beta\gamma}
=
\nabla_{\beta}\delta\Gamma^{\mu}_{\gamma\alpha}
-
\nabla_{\gamma}\delta\Gamma^{\mu}_{\beta\alpha},
\label{eq:BPalatini}
\end{equation}
and the antisymmetry of \(R^{\mu}{}_{\alpha\beta\gamma}\) in \(\beta\) and \(\gamma\), thus we get:
\begin{align}
2R_{\mu}{}^{\alpha\beta\gamma}\,\delta R^{\mu}{}_{\alpha\beta\gamma}
&=
2R_{\mu}{}^{\alpha\beta\gamma}
\left(
\nabla_{\beta}\delta\Gamma^{\mu}_{\gamma\alpha}
-
\nabla_{\gamma}\delta\Gamma^{\mu}_{\beta\alpha}
\right)
\nonumber\\
&=
4R_{\mu}{}^{\alpha\beta\gamma}\nabla_{\beta}\delta\Gamma^{\mu}_{\gamma\alpha}.
\label{eq:BdeltaR1}
\end{align}
After integration by parts,
\begin{equation}
2\int d^4x\,\sqrt{-g}\,
R_{\mu}{}^{\alpha\beta\gamma}\,\delta R^{\mu}{}_{\alpha\beta\gamma}
=
-4\int d^4x\,\sqrt{-g}\,
\left(\nabla_{\beta}R_{\mu}{}^{\alpha\beta\gamma}\right)\delta\Gamma^{\mu}_{\gamma\alpha}.
\label{eq:BdeltaR2}
\end{equation}
Now we use:
\begin{equation}
\delta\Gamma^{\mu}_{\alpha\beta}
=
\frac12 g^{\mu\lambda}
\left(
\nabla_{\alpha}\delta g_{\lambda\beta}
+
\nabla_{\beta}\delta g_{\lambda\alpha}
-
\nabla_{\lambda}\delta g_{\alpha\beta}
\right),
\label{eq:BdeltaGamma}
\end{equation}
and after integrating by parts once more and rearranging the terms, one finds:
\begin{equation}
2\int d^4x\,\sqrt{-g}\,
R_{\mu}{}^{\alpha\beta\gamma}\,\delta R^{\mu}{}_{\alpha\beta\gamma}
=
4\int d^4x\,\sqrt{-g}\,
\nabla^{\alpha}\nabla^{\beta}R_{\mu\alpha\nu\beta}\,\delta g^{\mu\nu}.
\label{eq:BdeltaR3}
\end{equation}
Hence,
\begin{equation}
\delta\!\left(\sqrt{-g}\,K\right)
=
\sqrt{-g}\,
\mathcal{K}_{\mu\nu}\,\delta g^{\mu\nu},
\label{eq:BdeltaKfinal}
\end{equation}
with
\begin{equation}
\mathcal{K}_{\mu\nu}
=
2R_{\mu}{}^{\alpha\beta\gamma}R_{\nu\alpha\beta\gamma}
-\frac12 g_{\mu\nu}K
+4\nabla^{\alpha}\nabla^{\beta}R_{\mu\alpha\beta\nu}.
\label{eq:BKmunu}
\end{equation}

The metric variation of the full action then gives:
\begin{equation}
\frac{\delta S}{\delta g^{\mu\nu}} = G_{\mu\nu}
+\alpha
\left(
2R_{\mu}{}^{\alpha\beta\gamma}R_{\nu\alpha\beta\gamma}
-\frac{1}{2}g_{\mu\nu}K
+4\nabla^\alpha\nabla^\beta R_{\mu\alpha\nu\beta}
\right),
\label{eq_Saction}
\end{equation}
where $G_{\mu\nu}$ is the standard Einstein equation (B2).

We now specialize Eq.~\eqref{eq_Saction} to the spatially flat FLRW metric in cosmic time (\ref{metrica}), with non-vanishing components (A5)-(A11) and $N(t)=1$.

The first Friedmann equation comes from the 00 component of (\ref{eq_Saction}). We have:
\begin{equation}
G_{00}=3\frac{\dot a^2}{a^2},
\label{eq:appB_einstein_00}
\end{equation}
\begin{equation}
2R_0{}^{\alpha\beta\gamma}R_{0\alpha\beta\gamma}
=
-12\frac{\ddot a^2}{a^2},
\label{eq:appB_term1_final}
\end{equation}
and
\begin{equation}
-\frac{1}{2}g_{00}K
=
6\left[
\left(\frac{\ddot a}{a}\right)^2
+
\left(\frac{\dot a}{a}\right)^4
\right].
\label{eq:appB_term2_final}
\end{equation}
The last and most delicate contribution, $
4\nabla^\alpha\nabla^\beta R_{0\alpha0\beta}$, can be treated by using the contracted Bianchi identity in the form:
\begin{equation}
\nabla^\beta R_{\mu\alpha\nu\beta}
=
\nabla_\alpha R_{\mu\nu}
-
\nabla_\mu R_{\alpha\nu}.
\label{eq:appB_contracted_bianchi}
\end{equation}
After some algebra and with the intermediary results:
\begin{equation}
\nabla_iR_{0i}
=
2\dot a\,\ddot a-2\frac{\dot a^3}{a}\,, \hspace{1cm}
\nabla_0R_{ii}
=
a\,\dddot{a}\,+3\dot a\,\ddot a-4\frac{\dot a^3}{a},
\label{eq:appB_nabla0Rii}
\end{equation}
we obtain:
\begin{equation}
4\nabla^\alpha\nabla^\beta R_{0\alpha0\beta}
=
12\left(
\frac{\dot a\,\dddot{a}\,}{a^2}
+
\frac{\dot a^2\ddot a}{a^3}
-
2\frac{\dot a^4}{a^4}
\right).
\label{eq:appB_4D00_final}
\end{equation}

Finally, substituting Eqs. (B13), (B14), (B15) and (B18) into the $00$ component of Eq.~\eqref{eq_Saction}, we obtain:
\begin{equation}
\frac{\dot a^2}{a^2}
-
2\alpha
\left(
\frac{\ddot a^2}{a^2}
-
2\frac{\ddot a\,\dot a^2}{a^3}
-
2\frac{\dddot{a}\,\dot a}{a^2}
+
3\frac{\dot a^4}{a^4}
\right)
=
0\,,
\label{eq:appB_friedmann1_final}
\end{equation}
which is precisely the first Friedmann equation obtained in Appendix A.

For the spatial components $ij$ the procedure is similar, with:
\begin{equation}
G_{ij}=-(\dot a^2+2a\ddot a)\delta_{ij},
\label{eq:appB_Gii_Tii}
\end{equation}
\begin{equation}
2R_i{}^{\alpha\beta\gamma}R_{j\alpha\beta\gamma}
=
\bigg(4\ddot a^2+8\frac{\dot a^4}{a^2}\bigg)\delta_{ij},
\label{eq:appB_term1_ii}
\end{equation}
\begin{equation}
-\frac{1}{2}g_{ij}K
=
\bigg(-6\ddot a^2
-6\frac{\dot a^4}{a^2}\bigg)\delta_{ij},
\label{eq:appB_term2_ii}
\end{equation}
\begin{equation}
4\nabla^\alpha\nabla^\beta R_{i\alpha j\beta}
=
\bigg(-4a\,\ddddot{a}\,
-8\dot a\,\dddot{a}\,
-4\ddot a^2
+24\frac{\dot a^2\ddot a}{a}
-8\frac{\dot a^4}{a^2}\bigg)\delta_{ij},
\label{eq:appB_4Dii_final}
\end{equation}
Finally, combining Eqs. (B20), (B21), (B22) and \eqref{eq:appB_4Dii_final} into (B12), we obtain:
\begin{equation}
\frac{\dot a^2}{a^2}
+
2\frac{\ddot a}{a}
+
2\alpha
\left(
3\frac{\ddot a^2}{a^2}
+
4\frac{\dot a\,\dddot{a}\,}{a^2}
+
2\frac{\ddddot{a}\,}{a}
-
12\frac{\dot a^2\ddot a}{a^3}
+
3\frac{\dot a^4}{a^4}
\right)
=0.
\label{eq:appB_second_friedmann_final}
\end{equation}

Equation~\eqref{eq:appB_second_friedmann_final} is the second Friedmann equation and completes the covariant verification of the cosmological system obtained in Appendix A.

\begin{acknowledgments}
This study was financed by the Coordenação de Aperfeiçoamento de Pessoal de Nível Superior - Brasil (CAPES) - Finance Code 001. JFJ acknowledges financial support from Conselho Nacional de Desenvolvimento Científico e Tecnológico (CNPq) (No. 314028/2023-4). SHP acknowledges financial support from Conselho Nacional de Desenvolvimento Científico e Tecnológico (CNPq) (No. 308469/2021 and 301775/2025-7).
\end{acknowledgments}


\end{document}